\documentclass[conference]{IEEEtran}
\IEEEoverridecommandlockouts

\usepackage{cite}
\usepackage{amsmath,amssymb,amsfonts,amsthm}
\usepackage{algorithmic}
\usepackage{graphicx}
\usepackage{textcomp}
\usepackage{xcolor}
\def\BibTeX{{\rm B\kern-.05em{\sc i\kern-.025em b}\kern-.08em
    T\kern-.1667em\lower.7ex\hbox{E}\kern-.125emX}}

\ifCLASSINFOpdf
\else
\fi
\usepackage{soul}
\usepackage{subfigure}
\usepackage{tabularx,booktabs}
\usepackage{url}
\usepackage{multirow}
\usepackage{boldline}
\usepackage{pifont}
\theoremstyle{definition}

\begin{document}

\title{On the Performance of Blockchain-enabled RAN-as-a-service in Beyond 5G Networks
%Performance Analysis of Blockchain-enabled as-a-service Spectrum Sharing in Future Networks
\thanks{This work was funded by the IN CERCA grant from the Secretaria d'Universitats i Recerca del departament d'Empresa i Coneixement de la
Generalitat de Catalunya, and partially from the Spanish MINECO grant TEC2017-88373-R (5G-REFINE) and Generalitat de Catalunya grant 2017 SGR 1195.}
}

\author{\IEEEauthorblockN{Francesc Wilhelmi}
\IEEEauthorblockA{Centre Tecnol\`ogic de Telecomunicacions de Catalunya \\
(CTTC/CERCA)\\
Castelldefels, Barcelona, Spain \\
fwilhelmi@cttc.cat}
\and
\IEEEauthorblockN{Lorenza Giupponi}
\IEEEauthorblockA{Centre Tecnol\`ogic de Telecomunicacions de Catalunya \\
(CTTC/CERCA)\\
Castelldefels, Barcelona, Spain \\
lorenza.giupponi@cttc.es}
}

\maketitle

\begin{abstract}
Blockchain (BC) technology can revolutionize the future of communications by enabling decentralized and open sharing networks. In this paper, we propose the application of BC to facilitate Mobile Network Operators (MNOs) and other players such as Verticals or Over-The-Top (OTT) service providers to exchange Radio Access Network (RAN) resources (e.g., infrastructure, spectrum) in a secure, flexible and autonomous manner. In particular, we propose a BC-enabled reverse auction mechanism for RAN sharing and dynamic users' service provision in Beyond 5G networks, and we analyze its potential advantages with respect to current service provisioning and RAN sharing schemes. Moreover, we study the delay and overheads incurred by the BC in the whole process, when running over both wireless and wired interfaces.
\end{abstract}


\IEEEpeerreviewmaketitle

%%%%%%%%%%%%%%%%%%%%%%%%%%%%
%% INTRODUCTION
%%%%%%%%%%%%%%%%%%%%%%%%%%%%
\section{Introduction}
From the third-generation (3G) of mobile communications~\cite{frisanco2008infrastructure}\cite{samdanis2016network}, network sharing has become a feasible, cost-effective solution to improve communication networks' capabilities (e.g., coverage, capacity) by splitting deployment and operational costs among multiple players such as mobile network operators (MNOs), virtual MNOs (MVNOs), or over-the-top (OTT) providers.
%As of today, with the advent of network virtualization~\cite{samdanis2016network} in the context 5G, the potential of network sharing is unleashed by allowing to leverage the existing infrastructure and spectrum resources to the highest extent.
In particular, cloudification, network function virtualization (NFV), and network slicing concepts are transforming communications systems as we know them. These concepts are triggering the emergence of a new sharing economy, aimed at limiting capital and operational expenditure (CAPEX/OPEX). 

A segment of the network which is receiving significant attention for network sharing is the radio access network (RAN), which agglutinates around 70\% of network management costs~\cite{gsma2019era}. The new Open-RAN (O-RAN)~\cite{yang2013openran} initiative also motivates the proliferation of RAN sharing and slicing solutions, so that the traditional concept of owning and controlling the network infrastructure becomes obsolete. This paves the way to a more sustainable paradigm in mobile communications, which favors the entry of new players to enrich the market, leading to enhanced services and infrastructure. An example is the emergence of new MVNOs delivering specialized services to increase niche markets, e.g., Google Fi, which offers cost-effective phone plans with additional services such as cloud storage, or discounts in Google-brand smartphones. 

% By deconstructing the functionalities carried out at a BS, it is possible to gather all the baseband processing to a central place

%Looking at beyond 5G (B5G) and 6G networks, sustainable network management strategies will be required to enable dynamic and intelligent sharing environments, where a huge variety of users’ relationships are available. 
In this context, we identify Blockchain (BC) technology as a key enabler for autonomous and extremely dynamic sustainable RAN/network sharing strategies in beyond 5G (B5G) and 6G networks. BC is particularly useful to provide trust and validity to the procedures of selling/buying resources among multiple stakeholders (e.g., an OTT provider requesting network resources to deliver a service to a third-party user), thanks to immutability, transparency, and security properties that it provides. BC can dramatically reduce the current time and effort required to define agreements among parties for sharing resources, based on legal contracts, and on the involvement of third-party intermediaries (e.g., banks, credit rating agencies, regulatory entities). These static and administratively heavy approaches prevent innovation, automation, and efficient, sustainable use of resources.

We propose BC to replace the traditional procedures for RAN sharing, with an automated economically-driven RAN management where operators can dynamically sublease their spectrum/RAN resources to other players for delivering services to users, increasing their coverage area, or improving their capacity. Smart contracts allow to automatically define and enforce SLAs (Service Level Agreements) among parties. Besides the automated RAN sharing, we also foresee a scenario where users advertise the need for data services through a BC, thus removing the need to be bound to a given MNO by a fixed contract. %Fig.~\ref{fig:blockchain_ecosystem} provides an overview of the BC ecosystem, where transactions and blocks are propagated through either dedicated/shared wireless or wired interfaces. Transactions and blocks are transmitted over a peer network, where peer nodes (or miners) keep the distributed ledger up to date.
%\begin{figure}[ht!]
%\includegraphics[width=\linewidth]{blockchain_ecosystem.pdf}
%\caption{Blockchain-enabled as-a-service network ecosystem.}
%\label{fig:blockchain_ecosystem}
%\end{figure}

Despite the benefits brought by a BC-enabled communication system, a large number of practical and implementation issues remain open~\cite{reyna2018blockchain}. In particular, the mining process associated to BC operation has been widely demonstrated to be very sensitive to the delays between peer nodes~\cite{gobel2016bitcoin}. For that reason, BC has been typically adopted through high-capacity networks, thus ignoring the potential impact of the communication latency. In particular, little work has been devoted to investigating the performance of a BC when ran over wireless networks, where wireless links may compromise the performance of a BC implementation by inflicting additional latency into the consensus mechanisms. A prominent example can be found in IEEE 802.11 networks, where the delay increases dramatically with the number of contending devices.

In this regard, the main contribution of this paper is twofold:
\begin{enumerate}
    \item First, we study the performance of the enabling BC network when realized through either wireless and wired links, in order to illustrate the viability of the BC application in dense, challenging wireless deployments. Decentralized consensus-based applications on which BC relies are heavily dependent on the propagation delay, which makes them vulnerable to undesired issues such as forks, double-spending, or re-transmissions. %For that, we put the realization of BC in wireless networks into question.
    \item Second, we foresee a BC-enabled RAN-as-a-service scenario for enabling transactions dynamically in future communications. %like the one presented in Fig.~\ref{fig:blockchain_ecosystem}
    In particular, we propose a hybrid blockchain scheme where multiple players (e.g., users, MNOs) acquire and provide network resources. Through event-driven simulations, we analyze the gains of the proposed scheme in terms of users' capacity and service acceptance in a variety of deployments that characterize multiple interactions among operators. With this, we aim to shed light on the feasibility of operating future networks automatically with BC technology.
    
\end{enumerate}

\section{Related Work}
\label{section:related_work}

BC technology establishes trust mechanisms that can remove the need for costly intermediaries and enable unprecedented levels of transparency and information sharing. As widely known, BC had a great impact on cryptocurrencies~\cite{nakamoto2019bitcoin}, but recently it also emerged as a powerful tool to revolutionize communications systems and to provide high savings in mobile networks. In~\cite{di2017smart}, the authors proposed to adopt smart contracts as a scalable tool to automatically generate and enforce SLAs, thus removing potential legal constraints in decentralized environments. This vision facilitates that new actors may appear as potential service providers, and new kinds of relationships can be established in an open sharing environment, shifting away from the traditional MNO paradigm with static resources and sharing mechanisms. Other areas of applications include privacy-preserving transactions~\cite{xu2021ran}, resource sharing~\cite{maksymyuk2020blockchain, xu2020blockchain}, or service provisioning~\cite{ling2019blockchain}, to name a few. 

Authors in \cite{maksymyuk2020blockchain} propose the usage of BC technology in 6G networks to simplify the interactions among spectrum regulators, MNOs, and end-users, defining some initial procedures and protocols required in a BC-empowered 6G ecosystem. 
Similarly, \cite{xu2020blockchain} overviews the benefits of introducing BC in spectrum management, and provides a set of illustrative scenarios where BC is relevant. More recently, a new {O-RAN}-based, BC-enabled architecture has been proposed in~\cite{xu2021ran} to empower P2P communications in 5G and beyond networks. By providing authentication and billing functionalities, BC has been shown to enrich RAN operation, thus potentially leading to lower latency and  privacy for applications such as Industrial IoT (IIoT) or autonomous driving. In~\cite{backman2017blockchain}, the authors propose a BC realization of the 5G network slice broker~\cite{samdanis2016network} for the future factoring use case. The concept of BC broker has also been discussed in~\cite{nour2019blockchain}, which was shown to be a key element to enable secure and reliable interactions among service and resource providers.

Closer in spirit to our work, we find the BC-enabled radio access framework proposed in~\cite{ling2019blockchain}. In such a framework, BC favors openness in a market whereby users can dynamically purchase digitized spectrum assets from Access Points (APs). Unlike the work in~\cite{ling2019blockchain}, we provide a broader view by considering a RAN ecosystem in which operators can exchange network resources as well, according to the varying user requirements. In particular, we propose using a hybrid BC system to adapt to the different types of transactions to be held in the envisioned ecosystem. The usage of public (or \textit{permissionless}) ledgers typically fits to highly decentralized applications, whereas private (or \textit{permissioned}) ones allow carrying out cost-effective, low-latency operations. The hybrid BC approach has been previously considered in~\cite{fan2020hybrid} for empowering Federated Learning (FL) in edge computing. %On one side, the transactions related to end-users are proposed to be maintained by a public blockchain, thus allowing any individual to request and provide services. Furthermore, the agreements among operators and other stakeholders are managed by a private (or consortium) blockchain.

%Despite the benefits projected by the previous work on BC-enabled communications systems, a large number of issues remain open, especially regarding practical implementation aspects~\cite{reyna2018blockchain}. The fact is that the mining, consensus-based operations associated with BC applications have been shown to be very sensitive to the delays among peer nodes and miners~\cite{gobel2016bitcoin}. For that reason, BC has been typically adopted through high-capacity networks. Nevertheless, to promote ubiquitous BC mechanisms, we identify wireless technologies as key. To the best of our knowledge, a little work has been done on analyzing the performance of BC when ran over wireless networks. For that reason, in this work, we evaluate the performance of the proposed BC-enabled as-a-service vision over contention-based IEEE 802.11 links.

%%%%%%%%%%%%%%%%%%%%%%%%%%%%
%% BC SOLUTION
%%%%%%%%%%%%%%%%%%%%%%%%%%%%
\section{Blockchain-enabled Dynamic RAN Sharing}
\label{section:solution_proposal}

%As for auctioning mechanisms, they have been widely applied in problems related to infrastructure sharing~\cite{franco2019brain}, Federated Learning (FL) in Internet-of-Things (IoT)~\cite{fan2020hybrid}, spectrum leasing~\cite{chen2017auction}, or Wi-Fi offloading~\cite{zhou2020draim,wang2017auction}. Auctioning and blockchain complement each other very well since smart contracts allow enforcing the rules of a given bidding process automatically. In this regard, the work in~\cite{koirala2019supply} provided an implementation of a BC-enabled reverse-auction bidding procedure for supply chains. 
%Moreover, we propose automatic auction procedures enabled by smart contracts to carry out transactions among users and operators or other providers. 

We focus on a scenario where two resource trading procedures are enabled. On the one hand, operators exchange radio and infrastructure resources dynamically, based on the varying users' demands. On the other hand, User Equipments (UEs) are not necessarily bound to a specific operator in the long term, but at any moment, can trigger requests for specific services. The proposed RAN-as-a-service vision is based on automated and auditable reverse-auction mechanisms, enabled by the BC and smart contract technologies. 
%Smart contracts facilitate an automatic, and auditable rational reverse auction mechanism among operators and service providers, interested in offering service. 
We propose an architecture based on two separate, but connected BCs, supporting the aforementioned functionalities:
\begin{itemize}
    \item \textbf{Service BC (public):} the service BC is used to record spontaneous service requests made by UEs, which are not associated to a fixed MNO through an already established contract. In addition, the BC registers the corresponding service delivery, provided by the selected operator or service provider, after a reverse auction mechanism is resolved. The service request is implemented through a smart contract defining the requested Key Performance Indicators (KPIs) for the required service, to be established in a SLA, setting the basis for service enforcement. This BC is public for the sake of openness, transparency, and decentralization.
    \item \textbf{RAN BC (private):} the RAN BC allows MNOs and other players to exchange RAN and network and radio resources such as spectrum or infrastructure. This BC is private and managed by a central auction authority, which fosters privacy and inhibits potential dishonest behaviors among operators.
\end{itemize}

\begin{figure}[ht!]
\includegraphics[width=\linewidth]{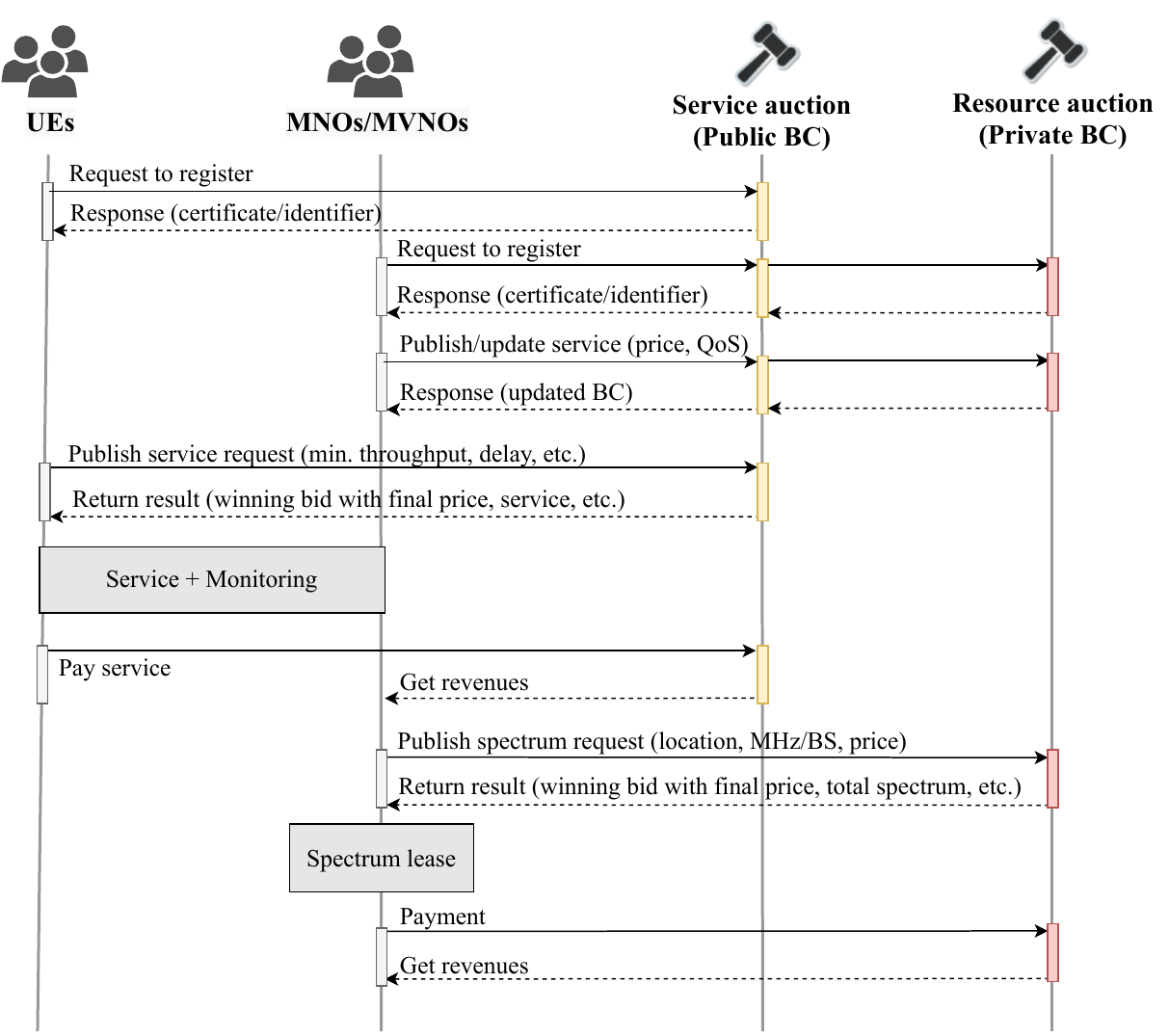}
\caption{RAN reverse-auction procedures enabled by public and private blockchains.}
\label{fig:ran_sharing_ecosystem}
\end{figure}

Fig.~\ref{fig:ran_sharing_ecosystem} shows the proposed BC-enabled auction mechanisms. On one side, UEs publish service requests to the public service BC, so that the reverse-auction is initiated and operators can place bids offering services. Upon completing the reverse-auction, the most satisfying bid is selected and the winner is automatically enforced to deliver the service by a smart contract defining the agreed SLA. The service is provided, monitored, and closed. Finally, payment is executed. As for the private BC, this allows operators or service providers in need for more RAN resources due to the extremely dynamic users' demand, to exchange infrastructure resources to increase coverage or capacity. Similar to the service auction, operators submit resource requests through the private BC so that a reverse-auction procedure is carried out to allow other operators to lease unused resources. %In both cases, the blockchain allows recording the requests and bids made by either UEs or operators, as well as the transactions associated to the resolution of the auction procedures.

%%%%%%%%%%%%%%%%%%%%%%%%%%%%
%% SYSTEM MODEL
%%%%%%%%%%%%%%%%%%%%%%%%%%%%
\section{System Model}
\label{section:system_model}
\subsection{Service auction}
Let us consider a set of $N$ UEs $\mathcal{U} = \{u_1,...,u_N\}$ with variable service demands $D_t = \{d_{1,t},...,d_{N,t}\}$ that subscribe dynamically to $M$ operators $\mathcal{O} = \{o_1,...,o_M\}$ through a reverse-auction procedure. The winner of a reverse-auction, i.e., the operator to which the user subscribes temporarily, can be selected based on multiple parameters (e.g., offered price, QoS, reputation) so that a given buyer's utility function is maximized:
\begin{equation}
    U^B_b = \max_{s \in S} \sum_{k=1}^K w_{b,k} x_{s,k},
    \label{eq:utility_buyer}
\end{equation}
where $w_{b,k} \in [0,1]$ is the weight that buyer $b$ assigns to feature $k$, and $x_{s,k}$ is a scalar value representing feature $k$ of seller $s$. In particular, given that the focus of this paper is not on the operator selection procedure, we consider that any user subscribes randomly to one of the available operators, which defines a random price between 0 and 1. This is useful to abstract the diversity on the criteria of multiple users, some of which may be intangible.
 
As for performance evaluation, we compute the capacity experienced by a given UE $u_n$ as a function of the signal-to-interference-plus-noise ratio (SINR) and the spectrum $b_n \in R^m(t)$ allocated by the selected operator $o_m$. As a result, the capacity experienced by user $u_n$ is:
\begin{equation}
    C_n = b_n \log_2(1 + \text{SINR}_n)
\end{equation}

Besides the users' capacity, we define the service acceptance~\cite{giupponi2009fuzzy} as a satisfaction metric, which depends on the service experienced by the user (in terms of bandwidth allocated) and the price $p$ paid for such a service. In particular, the service acceptance of user $u_n$ is defined as:
\begin{equation}
    A_n(t) = 1-\exp{(-C \cdot b_{n}^{\psi} \cdot p^\xi)}
    \label{eq:acceptance}
\end{equation}
where $C$ is a normalizing constant, and $\psi$ and $\xi$ are users' sensitivity to allocated bandwidth and price, respectively. As done in \cite{giupponi2009fuzzy}, we consider different $\psi$ and $\xi$ values to capture heterogeneity in terms of user profiles. %$C = 0.05$

\subsection{RAN auction}
Operators participate in both BC-enabled reverse-auction procedures to either deliver services to UEs or acquire/sublease RAN resources from/to other operators. Each operator owns a set of resources $R = \{r_{1},...,r_{M}\}$, which can be discretely split for sharing purposes, so that $r_i = \{r_{i,1},...,r_{i,S}\}$. To model the exchange of RAN resources, we define the temporary ownership of resources as $R(t) \rightarrow r_{i,s}^m(t)$, indicating that operator $o_m$ has the right of use of resource $r_{i,s}$ at time $t$. 
Through the proposed BC-enabled reverse-auction procedure, operators can automatically sublease the resources required by other operators at a fixed price, provided that the requested resources are unused. Defining variable prices set by operators is out of the scope of this paper, as it may lead to game theoretic implications. The price of network resources need to be carefully defined according to operators' expenses (e.g., equipment acquisition, site buildout, spectrum licenses, resources leasing, energy consumption, maintenance)~\cite{wisely2018capacity}.%in order to reach welfare-maximizing strategies

\subsection{BC's transaction confirmation delay}
To run the abovementioned auction procedures, we implement two separate BCs, namely a public BC (for keeping a record of service transactions) and a consortium BC (devoted to RAN infrastructure leasing transactions). In order to quantify the transaction confirmation delay of the BC-enabled dynamic RAN sharing mechanism, for both the BCs, we identify the following steps:
\begin{enumerate}
    \item First, UEs or operators upload transactions to the peer network ($T_\text{up}$). The incurred delay depends on the utilized communications technology (e.g., IEEE 802.11, 5G NR). 
    \item As transactions are collected by miners, they are used to fill blocks upon reaching the maximum block size $S_B$, or the expiration of timer $T_\text{wait}$. The rate at which transactions arrive, together with the timer value, leads to a queuing time $T_\text{queue}$, which has been previously analytically modelled in~\cite{wilhelmi2021discrete}.
    \item Once a block is ready to be mined, peer nodes execute a consensus algorithm to find the block's nonce. As widely considered in the literature~\cite{decker2013information, biais2019blockchain}, the mining (or block generation) time, $T_\text{mine}$, can be modeled with an exponential random variable with parameter $\mu$.
    \item When a block is mined, it is propagated ($T_\text{prop}$) and eventually appended to the BC upon successful block propagation.
\end{enumerate}

It is important to notice that, because of the nature of the distributed consensus mechanism, a BC network is susceptible to experience forks. Forks can be seen as a lack of agreement among the peer nodes that maintain a distributed ledger, which may lead to invalidating transactions and blocks. The block propagation delay is of crucial importance due to its direct relationship with the probability of fork generation. The existence of forks may have severe consequences in the performance of the BC since they contribute to increasing the instability and the confirmation delay of the system.

The transaction confirmation delay fpr each of the proposed BCs can be approximated as follows:
\begin{equation}
    T_\text{c} = T_\text{up} + \frac{(T_\text{queue} + T_\text{mine} + T_\text{prop})}{1-p_\text{fork}},
  %  \nonumber
\end{equation}
where $p_\text{fork}$ is the fork probability. For the details of the derivation to compute $T_\text{c}$, the reader is referred to the previous work in~\cite{wilhelmi2021discrete}.

%In both BCs, peers (miners) fill blocks with transactions upon reaching the maximum block size $S_B$, or when a timer $T_\text{w}$ expires, thus leading to a queuing time $T_\text{q}$. Once a block is ready to be mined, peer nodes execute a consensus algorithm to find the block's nonce. As widely considered in the literature~\cite{decker2013information, biais2019blockchain}, the mining (or block generation) time, $T_\text{mine}$, can be modeled with an exponential random variable with parameter $\mu$. When a block is mined, it is propagated and eventually appended to the blockchain upon successful block propagation. The block propagation delay is of crucial importance due to its direct relationship with forks, which occur as a result of the consensus procedure. The existence of forks may have severe consequences in the performance of the BC since they contribute to increasing the end-to-end latency of the system.

%%%%%%%%%%%%%%%%%%%%%%%%%%%%
%% PERFORMANCE EVALUATION
%%%%%%%%%%%%%%%%%%%%%%%%%%%%
\section{Numerical Results}
\label{section:results}

In this section, we first study the main factors that affect the BC operation through simulations. In particular, we focus on the performance of the proposed BC-enabled reverse-auction procedure when supported by contention-based wireless links and backhaul interfaces. With that, we aim at showcasing the potential benefits and drawbacks of adopting such technology in communications. Then, we demonstrate the gains of the proposed BC-enabled sharing scheme in terms of the capacity and service acceptance experienced by users. Our simulation-based analysis is conducted in a typical hexagonal cellular deployment, with 19 sectors
%as shown in Figure~\ref{fig:random_deployment}, 
where UEs are randomly dropped. To upload transactions and propagate blocks, the public BC operates through unlicensed IEEE 802.11 wireless links, whereas the private one makes use of the wired X2/Xn interfaces~\cite{assefa2017sdn}. Specifically, we consider a fixed latency of 10~ms for the consortium BC. The delay in the public BC relies on IEEE 802.11 technology and depends on the status of the wireless channel and the set of overlapping nodes (see Appendix~\ref{section:delay_wifi}). The simulation parameters used throughout this section are collected in Table~\ref{tbl:simulation_parameters}. 

%\begin{figure}[ht!]
%\centering
%\includegraphics[width=.8\linewidth]{random_deployment.png}
%\caption{Random cellular-based deployment. \textcolor{red}{Try to make this figure nicer.}}
%\label{fig:random_deployment}
%\end{figure}

% \usepackage{graphicx}
\begin{table}[]
\centering
\caption{Simulation parameters.}
\label{tbl:simulation_parameters}
\resizebox{\columnwidth}{!}{%
\begin{tabular}{|c|l|c|}
\hline
\textbf{Parameter} & \multicolumn{1}{c|}{\textbf{Description}} & \textbf{Value} \\ \hline
$A$ & Num. of APs/cells & 19 \\ \hline
$R$ & Cell radius & 10 m \\ \hline
$M$ & Number of MNOs/MVNOs & 2-5 \\ \hline
$N$ & Number of UEs & 200 \\ \hline
$B$ & Bandwidth & 20 MHz \\ \hline
$F_c$ & Carrier frequency & 5 GHz \\ \hline
$MCS$ & Modulation and coding scheme & 0-11 \\ \hline
$P_t$ & Transmit power & 20 dBm \\ \hline
$PL(n)$ & Loss at distance $d$ & $P_t-PL_0+10 \alpha \log_{10}(d) + \frac{\sigma}{2} + \frac{d}{10} \frac{\gamma}{2}$\\ \hline
$PL_0$ & Loss at the reference dist. & 5 dB \\ \hline
$\alpha$ & Path-loss exponent & 4.4 \\ \hline
$\sigma$ & Shadowing factor & 9.5 \\ \hline
$\gamma$ & Obstacles factor & 30 \\ \hline
$\mu$ & Mining capacity & 10 blocks/s \\ \hline
$L_T$ & Transaction length & 3.000 bits \\ \hline
$T_w$ & Maximum waiting time & 5 s \\ \hline
$\psi$ & Users' bandwidth sensitivity & \multicolumn{1}{c|}{0.01 to 0.2} \\ \hline
$\xi$ & Users' price sensitivity & \multicolumn{1}{c|}{0.01 to 0.2} \\ \hline
$\delta_{X2/Xn}$ & Latency X2/Xn interfaces & \multicolumn{1}{c|}{10 ms} \\ \hline
$\delta_{11ax}$ & Latency IEEE 802.11ax interfaces & \multicolumn{1}{c|}{See Appendix~\ref{section:delay_wifi}} \\ \hline
\end{tabular}%
}
\end{table}

\subsection{BC Performance}
\begin{figure}[ht!]
\centering
\subfigure[$T_\text{wait}=0.1$s]{\includegraphics[width=\columnwidth]{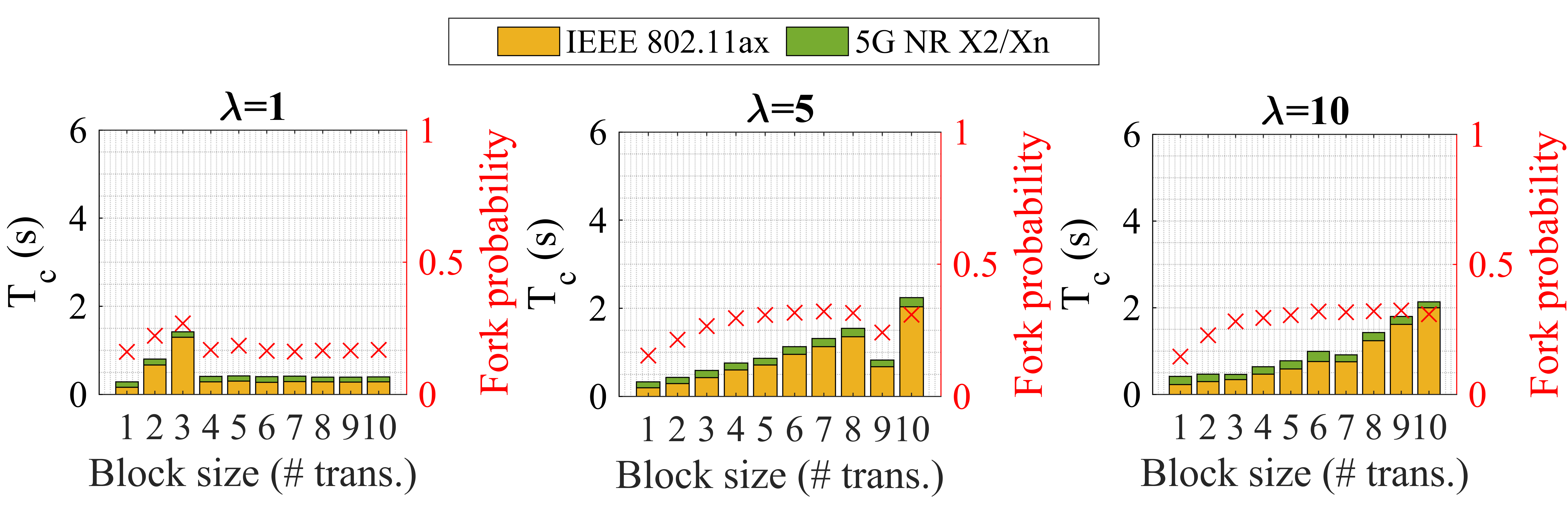}\label{delays_bc_1}}
\subfigure[$T_\text{wait}=5$s]{\includegraphics[width=\columnwidth]{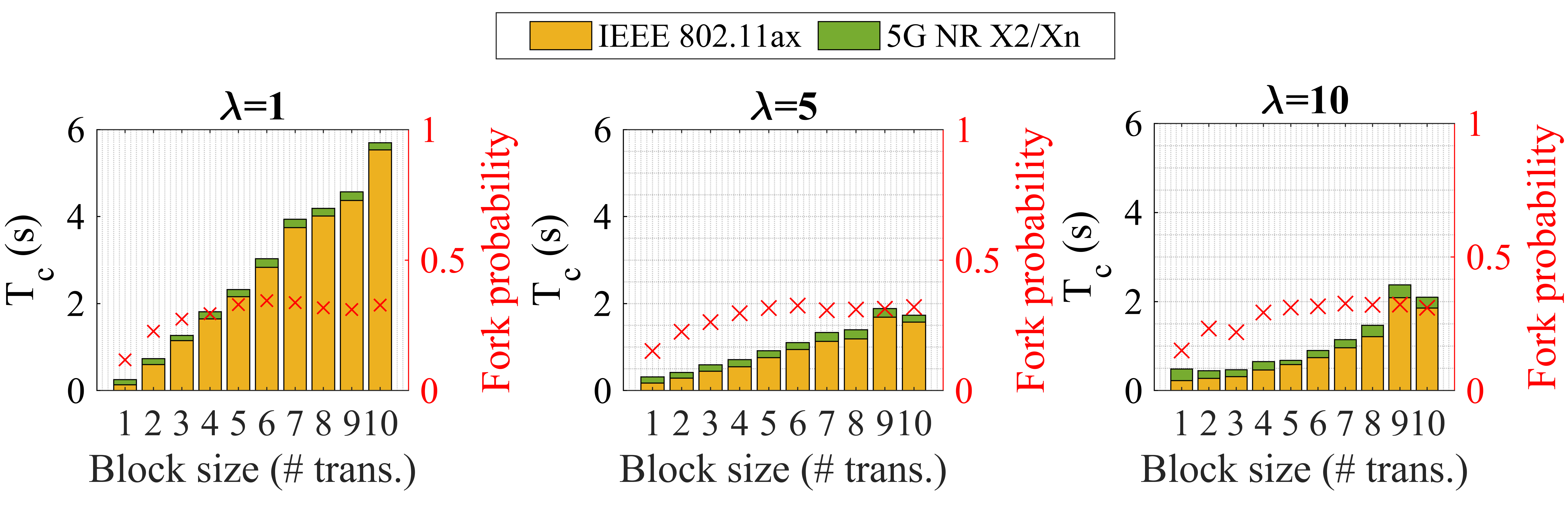}\label{delays_bc_2}} 
\caption{Transactions confirmation delay of the service (public) and RAN (private) BCs, for different timers ($T_\text{wait}$), block sizes ($S^B$) and user arrival rates ($\lambda$). The fork probability of the public BC is shown in red.}
\label{fig:delays_blockchain}
\end{figure}

\begin{figure}[ht!]
\centering
\includegraphics[width=\columnwidth]{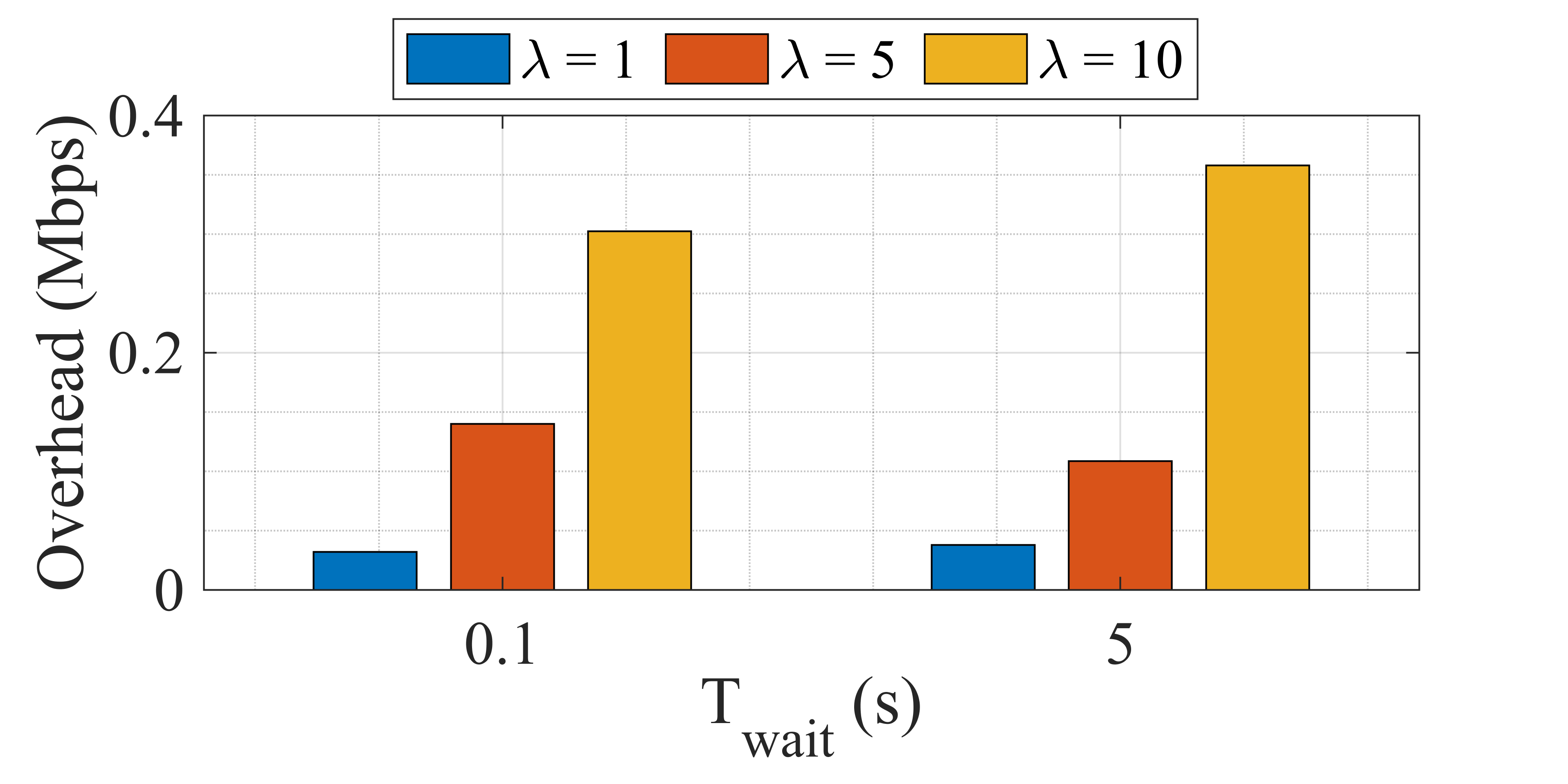}
\caption{Overhead incurred by the public BC, for different timers ($T_\text{wait}$), block sizes ($S^B$) and user arrival rates ($\lambda$).}
\label{fig:overhead_bc}
\end{figure}

The adoption of the proposed BC-enabled service provisioning and RAN sharing entails certain costs in terms of delay and overhead. In a BC, the confirmation delay is measured as the time it takes to validate transactions, including the mining procedure. These overheads are strongly tied to the underlying communication technology used by the BC. In particular, we study the cases of IEEE 802.11 (for the service blockchain) and 5G X2/Xn backhaul links (for the RAN BC). Fig.~\ref{fig:delays_blockchain} shows the confirmation delay experienced by each type of BC under different settings, while Fig.~\ref{fig:overhead_bc} depicts the associated communication overhead. Through these results, we analyze the effect of important BC parameters, including the block size, the maximum waiting time, and the system arrivals rate. It is important to notice that, due to the nature of the two BCs, the public one is susceptible to forks, while the private one is not. This can be seen as a direct consequence of the type of consensus mechanisms that can be applied to each of the BC, namely, for instance, Proof-of-Work (PoW) and Proof-of-Authority (PoA), respectively. 

As shown in Fig.~\ref{fig:delays_blockchain}, the block size ($S^B$) has a great impact on the confirmation delay, which tends to increase as more transactions are required for filling a block. This issue is exacerbated when the maximum waiting time ($T_\text{w}$) is high and the number of system arrivals ($\lambda$) is low. Therefore, setting a low waiting time can help decrease latency when the BC's load is low. However, increasing the throughput of the BC may lead to additional overhead, as depicted in Fig.~\ref{fig:overhead_bc}. In general, finding the optimal BC parameters is a complex task. For instance, increasing the block size is useful to reduce the BC overhead in terms of headers and mining time. However, it contributes to increasing the fork probability, thus potentially leading to additional confirmation delay and overhead.
% Queue model BC: \cite{wilhelmi2021discrete}

\subsection{Users' Performance}
\begin{figure}[ht!]
\centering
\subfigure[Capacity]{\includegraphics[width=\columnwidth]{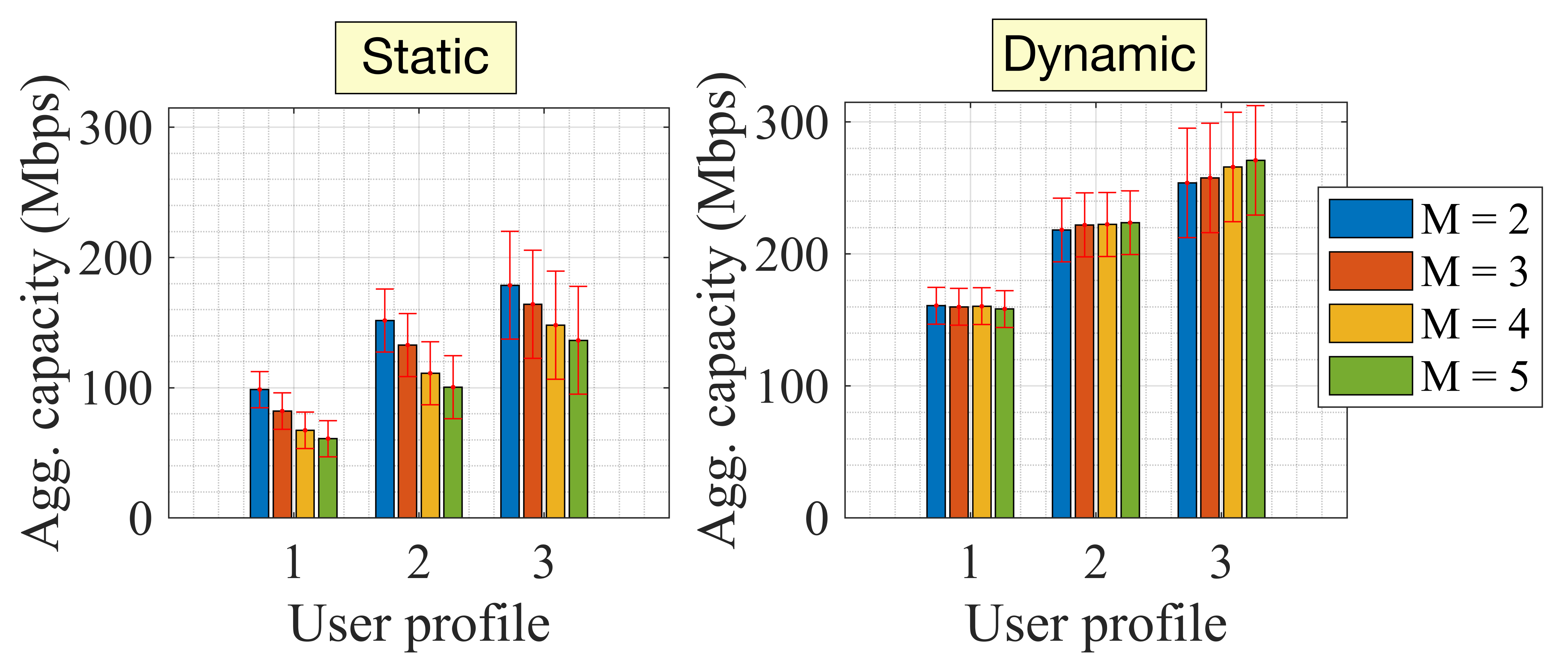}\label{mean-cap-aps}} 
\subfigure[Service acceptance]{\includegraphics[width=\columnwidth]{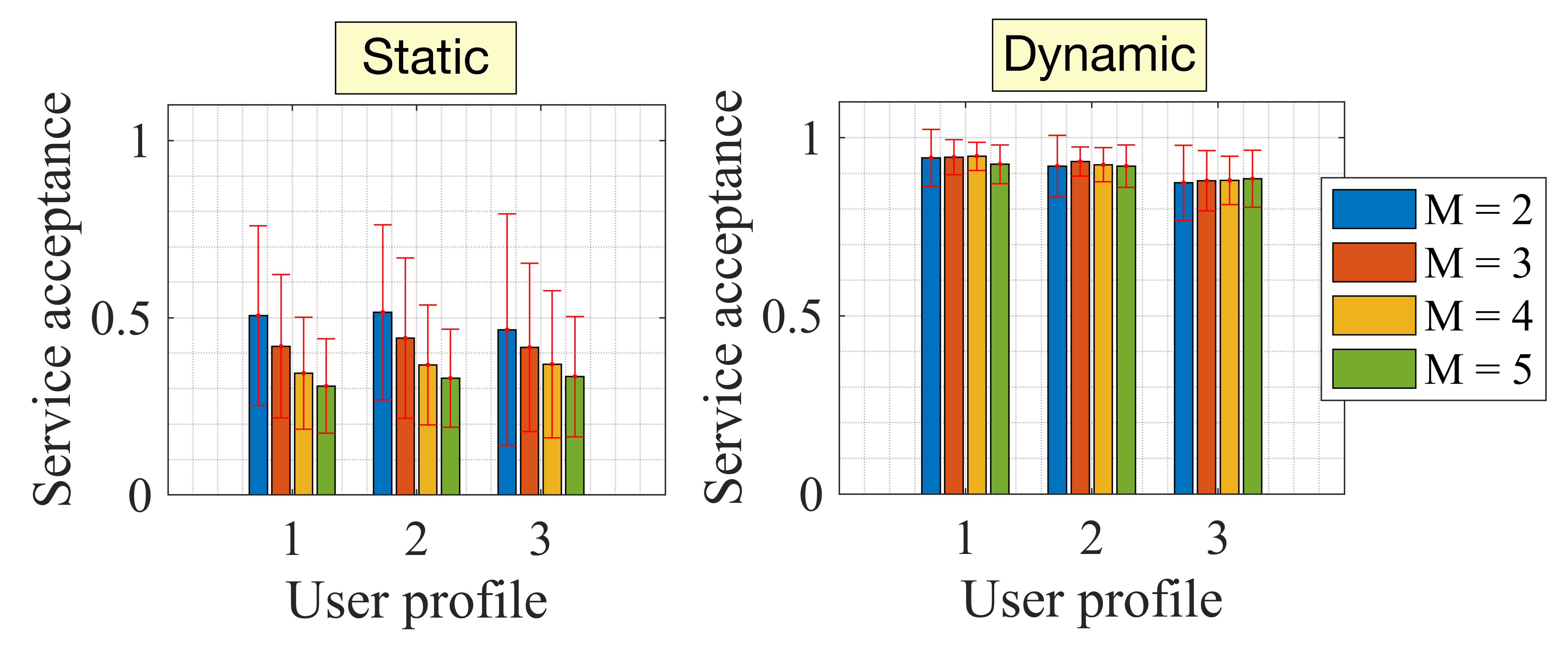}\label{mean-load-aps}} 
\caption{Mean aggregate capacity and service acceptance for the static (left) and dynamic (right) RAN sharing approaches in random deployments. Different number of user profiles and operators ($M$) are considered.}
\label{fig:performance_random_deployments}
\end{figure}

To demonstrate the superiority of the proposed BC approach for enabling dynamic resource sharing in future networks, in Fig.~\ref{fig:performance_random_deployments}, we plot the overall performance obtained in terms of aggregate capacity and service acceptance (see Eq.~\eqref{eq:acceptance}). We compare results obtained through BC-enabled RAN sharing, to a static approach which does not enable resource trading among operators. Random deployments with a varying number of operators, fixed available resources in the scenario, and different user profiles have been considered. We consider three different user profiles based on the AP share required by each user: 1) low traffic (0.001 to 0.01), 2) average traffic (0.05 to 0.02), 3) and high traffic (0.01 to 0.025). As for the BC configuration, we have considered $S^B = 15.000$ bits, and $T_w = 5$ s for both service and RAN BCs.

As illustrated by Fig.~\ref{fig:performance_random_deployments}, when RAN sharing is not enabled, both the capacity and service acceptance decrease with number of operators. This situation is properly addressed by the BC-enabled RAN sharing solution, which favors diversity through the exchange of resources among operators. With this, UEs experience improved data services. As for the different user profiles, the capacity improves as a result of the increasing traffic demands, which directly impacts the total load experienced by the different APs.

\subsection{A use case: A traditional MNO vs a Virtual MNO}

\begin{figure}[ht!]
\centering
\subfigure[\{1,0\} RAN ownership ratio]{\includegraphics[width=\columnwidth]{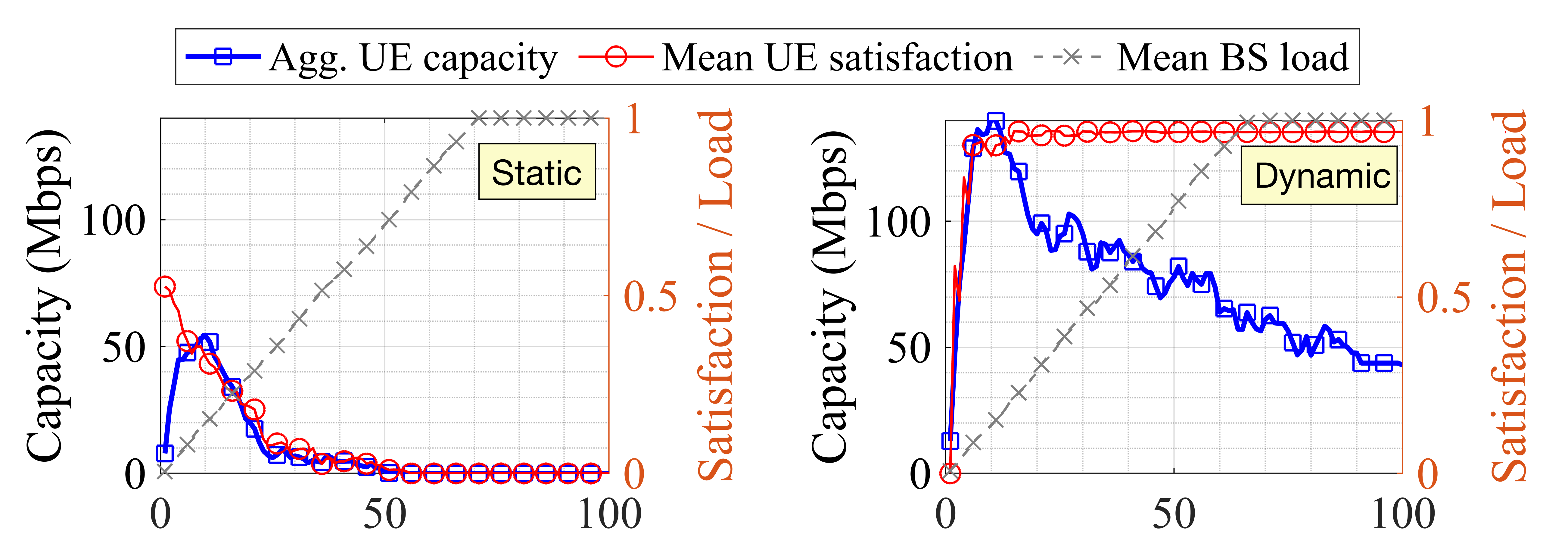}\label{two_operators_1}} 
\subfigure[\{0.5,0.5\} RAN ownership ratio]{\includegraphics[width=\columnwidth]{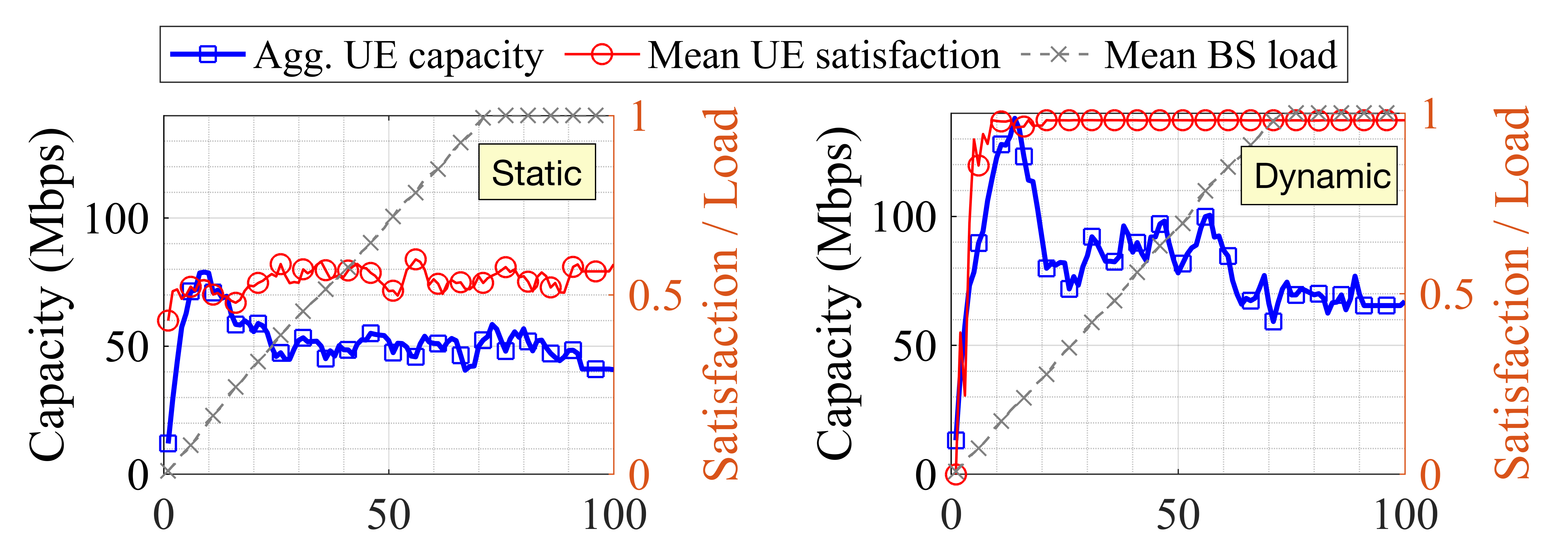}\label{two_operators_2}} 
\caption{Evolution of the aggregate capacity (in blue) and the mean satisfaction (in red) obtained by UEs in a two-operators scenario, for both static and dynamic RAN sharing approaches. The mean APs' load is shown in gray.}
\label{fig:temporal_capacity}
\end{figure}

To better illustrate the behavior of the sharing mechanism, we now propose a use case in which a traditional MNO may exchange resources with an incoming virtual operator. Fig.~\ref{fig:temporal_capacity} shows the temporal performance obtained in such a deployment for two different situations: 
\begin{enumerate}
    \item The MNO owns all the RAN resources (with ratio \{1,0\}), which can be shared if the BC-enabled RAN sharing procedure is enabled.
    \item The MNO and the MVNO sign a long-term agreement whereby resources are distributed with ratio \{0.5,0.5\}.
\end{enumerate}

As shown in Fig.~\ref{two_operators_1}, the static case leads to a very low capacity and service acceptance. The reason is that a single operator monopolizes all the resources, which leaves some users subscribed to the other operator without service. In the second case (see Fig.~\ref{two_operators_2}), the static RAN sharing improves the performance, in terms of both aggregated capacity and users' satisfaction. Nevertheless, the dynamic BC-enabled RAN sharing approach allows to better respond to the varying user requirements, which leads to an even higher system capacity and service acceptance. An important conclusion that can be drawn from this is that the current long-term static RAN sharing agreements lack the necessary flexibility in dense 5G/6G scenarios. To address this issue, BC emerges as a powerful solution to boost dynamism safely and reliably.

%%%%%%%%%%%%%%%%%%%%%%%%%%%%
%% CONCLUSIONS
%%%%%%%%%%%%%%%%%%%%%%%%%%%%
\section{Conclusions}
\label{section:conclusions}
In this paper, we have proposed to adopt BC technology to enable RAN-as-a-service for future 5G/6G networks. With BC, it is possible to establish a secure, open environment where the existing RAN infrastructure can be shared by different parties. Through simulation results, we have shown that the BC solution fosters competitiveness, which is useful to reduce and share operational and management costs in networks, thus allowing to improve the services offered to UEs. As for the performance of the BC, our simulation results have shown that BC technology is very susceptible to the communication delay for broadcasting transactions and blocks. Depending on the technology used to run a BC, the system performance can be drastically compromised. In particular, we have analyzed the behavior of BC when operated with IEEE 802.11 and 5G NR X2/Xn links. Our results have shown that links with constant delays (e.g., dedicated interfaces) favor BC's operation, whereas contention-based networks generate high instability when becoming congested, which leads to a high number of re-transmissions and forks.

\appendix

\subsection{Characterization of IEEE 802.11ax delays}
\label{section:delay_wifi}
We characterize the delay of IEEE 802.11 networks through the Bianchi's Distributed Coordination Function (DCF) model \cite{bianchi2000performance}, which defines the average duration of a generic slot $E[T_\text{slot}]$ in a saturated overlapping basic service set (OBSS) as:
\begin{equation}
    E[T_\text{slot}] = p_{\text{slot,e}} T_\text{slot,e} + p_{\text{slot,s}} T_\text{slot,s} + p_{\text{slot,c}} T_\text{slot,c},
\end{equation}
where the duration of each type of slot is given by:
\begin{equation}
  \resizebox{\columnwidth}{!}{%
   $\begin{cases}
       T_\text{slot,e} = 9 \mu s\\
       T_\text{slot,s} = T_{RTS} + 3\cdot T_{SIFS} + T_{CTS} + T_{DATA} + T_{ACK}\\
       T_\text{slot,c} = T_{RTS} + T_{DIFS}
   \end{cases} $
   }
\end{equation}

For $N$ overlapping nodes, the probability of each type of slot is calculated as follows:
\begin{equation}
    \begin{cases}
       p_{\text{slot,e}}=(1-\tau)^N\\
       p_{\text{slot,s}}=N\tau(1-\tau)^{N-1}\\
       p_{\text{slot,c}}=1-p_\text{slot,e}-p_\text{slot,s}
   \end{cases} 
\end{equation}

%Provided that a fixed contention window (CW) is applied, the probability of transmitting in a given slot $\tau$ is obtained from:
%\begin{equation}
%\tau=\frac{1}{E[\phi]+1}=\frac{2}{\text{CW}+1},
%\end{equation}
%where $E[\phi]=\frac{\text{CW}-1}{2}$. 
Where the expected backoff duration is given by:
\begin{equation}
   \text{E}[\phi] = \sum_{w=0}^{w_{max}} \pi_w \text{E}[\text{CW}(w)],
\end{equation}
where $\pi_w$ is the probability of being in stage $w$, so that $\text{CW}(w) = 2^w \text{CW}_{min}$. In particular, for a maximum number of re-transmissions $R_{max}$, the stage probability $\pi_w$ is:
\begin{equation}
\pi_w = \frac{p^w(1-p_c)}{1-p_c^{R_{max}+1}}
\end{equation}

\ifCLASSOPTIONcaptionsoff
\newpage
\fi

\bibliographystyle{IEEEtran}
\bibliography{bibliography}

\end{document}